\documentclass[a4paper,twoside,12pt]{article}
\usepackage{amsmath,amssymb}
\usepackage{verbatim}
\usepackage{bbm}
\usepackage{enumerate}
\usepackage{stmaryrd}

\DeclareMathOperator{\RE}{Re}
\DeclareMathOperator{\image}{image}

\newcommand{\dd}{\mathrm{d}}
\newcommand{\p}{\partial}

\renewcommand{\epsilon}{USE varepsilon INSTEAD}

\newcommand{\C}{\mathbbm{C}}

\newcommand{\R}{\mathbbm{R}}
\newcommand{\Z}{\mathbbm{Z}}

\newtheorem{defn}{Definition}{} 
\newtheorem{lemma}{Lemma}{} 

\newcommand{\SU}{\mathrm{SU}(2)}
\newcommand{\SL}{\mathrm{SL}(2,\C)}
\newcommand{\Ss}{{\boldsymbol\Pi}}
\newcommand{\Ub}{\mathbf{U}}
\newcommand{\Pb}{\widehat{\mathbf{P}}}

\newcommand{\Back}{L}
\newcommand{\Weyl}{\text{\bf Weyl}}
\newcommand{\GR}[1]{$\text{(#1)}_{\text{GR}}$}

\begin{document}

\def\thesubsection{\arabic{subsection}}


\noindent {\bf\Large A class of gauges for the Einstein equations}
\vskip 5mm
\noindent {\bf Michael Reiterer, Eugene Trubowitz}
\vskip 1mm
\noindent Department of Mathematics, ETH Zurich, Switzerland
\vskip 4mm
\noindent {\bf Abstract:} A class of gauges for the Einstein vacuum equations is introduced, along with three symmetric hyperbolic systems. The first implies the local realizability of the gauge. The second is the dynamical subset of the field equations. The third is used to show that the constraints propagate.\\
The gauges are for an orthonormal frame formalism, with first order, quadratically nonlinear equations.
The unknowns are 16 frame components and 28 connection components. After gauge-fixing, a total of 33 remain.
\vskip 8mm
\subsection{Introduction}
\subsubsection{The orthonormal frame formalism used in this paper}\label{sudskkdhfdk}
Introduce the Lie algebra $\Lambda = \R^4 \times (\R \times \mathrm{so}(1,3))$.
A quick overview of the formalism is obtained by counting the independent degrees of freedom of its constituents\footnote{Essentially: coordinate transformations + $(\R_+\times \mathrm{O}(1,3))$ frame transformations; frame + connection; torsion + curvature modulo Weyl. Torsion=0 is here a field equation!}. The number of
\vskip 2mm
\noindent \rule{30pt}{0pt}\begin{minipage}{12cm}
$\bullet$\;local gauge degrees of freedom\\
$\bullet$\;components of the unknown field\\
$\bullet$\;(first order, quadratically nonlinear) field equations
\end{minipage}
\vskip 2mm
\noindent
are equal, respectively, to 
\vskip 2mm
\noindent \rule{30pt}{0pt}\begin{minipage}{12cm}
$\bullet$\;$\dim_{\R}\,\Lambda = 4 + 7 = 11$\\
$\bullet$\;$\dim_{\R}\, (\R^4)^{\ast}\otimes_{\R}\Lambda = 16 + 28 = 44$\\
$\bullet$\;$\dim_{\R}\, [((\R^4)^{\ast} \wedge_{\R} (\R^4)^{\ast})\otimes_{\R} \Lambda]\,/\,\Weyl_{\R} = 24 + 42-10 = 56$
\end{minipage}
\vskip 2mm
\noindent
By the Einstein vacuum equations, the curvature vanishes only modulo elements of a subspace $\Weyl_{\R}\hskip-1pt\subset\hskip-1pt((\R^4)^{\ast} \wedge_{\R} (\R^4)^{\ast}) \otimes_{\R} \mathrm{so}(1,3)$, $\dim_{\R}\Weyl_{\R}\hskip-1pt=\hskip-1pt10$.
\vskip 2mm
\emph{The algebraic structure of the principal part\footnote{Principal part: the union of terms \emph{with first derivatives} of the unknown.\\
Non-principal part: the union of terms \emph{without derivatives} of the unknown.} of the equations is central to this paper.} This structure is already in the linear model problem (LMP)
\vskip 2mm
\noindent \rule{30pt}{0pt}\begin{minipage}{12cm}
$\bullet$\;$F \mapsto F + \dd \xi$ are the infinitesimal local gauge transformations\\
$\bullet$\;$U \in \Weyl_{\R}$ (pointwise) are the field equations; here $U = \dd F$\\
$\bullet$\;$\dd U = 0$ are differential identities for $U$
\end{minipage}
\vskip 2mm
\noindent
with $\Lambda$-valued differential forms $\xi$, $F$ (the unknown), $U$ of degree $0$, $1$, $2$. More details about this formalism are given as needed, in Section \ref{appgr}. 

\subsubsection{Gauge-fixing and symmetric hyperbolicity}
The gauges in this paper are fields with values in a vector space, on which $\SL$ acts.
These gauges impose $11$ linear conditions, that restrict the unknown field to a subspace with $\dim_{\R} = 33$,  see \eqref{eq:gaugefix} in Section \ref{lmpsec} and \eqref{eq:gcgr} in Subsection \ref{fddskhkfdshfdhfdk}. Accordingly, there are $33$ dynamical equations, and $56-33=23$ constraint equations.

These gauges are introduced with the explicit goal of obtaining the three
\emph{symmetric hyperbolic} systems\footnote{Symmetric hyperbolic systems were used, in this context, by H.\,Friedrich \cite{Fr}.}\,\footnote{See \cite{Tay} for the general theory of local existence and uniqueness for quasilinear symmetric hyperbolic systems, and finite speed of propagation.}\,\footnote{Example: The familiar $(\p_0 + i\,\mathrm{curl})\mathbf{v} = 0$ on $\R^4$ is a linear, constant coefficient, symmetric hyperbolic system. Here $\mathbf{v} = \mathbf{E} + i\mathbf{B}$ is a complex vector field, $\mathbf{v} = (v^1,v^2,v^3)$. In this example, symmetric hyperbolicity means that the $3\times 3$ matrix $\theta(\p_0 + i\,\mathrm{curl})$ is \emph{Hermitian for all} real one-forms $\theta = (\theta_0,\boldsymbol{\theta})$ and \emph{positive for some} such $\theta$. Here, it is positive iff $\theta$ is future timelike, $\theta_0 > |\boldsymbol{\theta}|_{\R^3}$.} that are mentioned in the abstract. Their dimensions are $11\times 11$ and $33\times 33$ and $23\times 23$.

Symmetric hyperbolicity is a property of \emph{the algebraic structure of just the principal part of the equations}. Thus, the  algebraic manipulations, used to derive symmetric hyperbolic systems, can be presented within the context of a linear model problem on $\R^4$ (LMP), in self-contained Sections \ref{sec:notpre} and \ref{lmpsec}.
Section \ref{appgr} explains, why the LMP results are results about general relativity.

\subsubsection{Comparison with the Newman-Penrose-Friedrich formalism}\label{npfform}
The Newman-Penrose \cite{NP} formalism differs from Subsection \ref{sudskkdhfdk} by the addition of the Weyl curvature to the list of unknowns, and the \emph{differential Bianchi equations} to the list of (still first order, quadratically nonlinear) field equations.

Friedrich \cite{Fr} derived differential identities for the field equations, and showed that by gauge-fixing, the field equations can be reduced to symmetric hyperbolic systems. Other examples of gauges that admit such a symmetric hyperbolic reduction have since been found. The essential strategy is to: gauge-fix frame and connection; derive `simple' equations\footnote{Here, `simple' refers to equations that are easily symmetric hyperbolic, e.g.\,with diagonal matrix differential operator, a `coupled system of ordinary differential equations'.} for frame and connection; bring the differential Bianchi equations for the Weyl curvature into symmetric hyperbolic form. One way or another, the last step uses the Bel-Robinson tensor. 

In this approach, gauge-fixing and the derivation of symmetric hyperbolic systems can be carried out almost independently: the first concerns only frame and connection, the second only the Weyl curvature\,/\,differential Bianchi equations. By contrast, in the present paper, gauge-fixing and the derivation of symmetric hyperbolic systems is correlated: both concern frame and connection.

\subsection{Notation and definitions}\label{sec:notpre}
\begin{enumerate}[$\bullet$]
\item $z \in \C$ has complex conjugate $\overline{z}$ and real part $\RE z = \tfrac{1}{2}(z+\overline{z})$.\\
Fields in this paper are complex. Reality conditions are spelled out.
\item Small Latin indices $a,b,\ldots$ take values in the (spinor) index set $\mathcal{S} = \{0,1\}$.
\item $\varepsilon^{ab}$, $\varepsilon_{ab}$ are defined by $\varepsilon^{ab} = -\varepsilon^{ba}$ and $\varepsilon_{ab} = -\varepsilon_{ba}$ and $\varepsilon^{01} = \varepsilon_{01} = 1$.\\
 $\mathbf{S}^{a_1\ldots a_k}_{b_1\ldots b_k} = (k!)^{-1} \sum_{\pi \in S_k} \delta^{a_1}_{b_{\pi(1)}}\cdots \delta^{a_k}_{b_{\pi(k)}}$, total symmetrization of $\mathcal{S}$-indices\footnote{Here, $S_k$ is the symmetric group on $\{1,\ldots,k\}$.}.\\
 Observe that $\varepsilon^{ka}\varepsilon_{kb} = \delta^a_b$ and $\varepsilon^{ak}\varepsilon_{kb} = -\delta^a_b$
and $\delta^i_a\delta^j_b = \mathbf{S}^{ij}_{ab} + \tfrac{1}{2}\varepsilon^{ij}\varepsilon_{ab}$.
\item Small Greek indices $\alpha,\beta,\ldots$ take values in the (coordinate) index set $\mathcal{C}$ with $|\mathcal{C}| = 4$.
The index set $\mathcal{C}$ is assumed to be disjoint from $\mathcal{S}$.
\item Capital Latin indices $A,B,\ldots$ take values in $\mathcal{C}\cup (\mathcal{S}\times \mathcal{S}\times \mathcal{S}\times \mathcal{S})$. Namely, $(f^A)$ is short for $(f^{\alpha}) \oplus ({f_{ab}}^{cd})$, and $(f_A)$ is short for\footnotemark\:$(f_{\alpha}) \oplus ({f^{ab}}_{cd})$.\footnotetext{Example of a summation that involves capital Latin indices:\\
$f^Kg^Lh_{iKL} = 
f^{\alpha}g^{\beta}h_{i\alpha \beta}
+ f^{\alpha} g_{mn}^{\phantom{mn}pq} h_{i \alpha \phantom{mn} pq}^{\phantom{i\alpha} mn}
+ f_{ab}^{\phantom{ab}ef} g^{\beta} h_{i\phantom{ab} ef \beta }^{\phantom{i}ab}
+ f_{ab}^{\phantom{ab}ef} g_{mn}^{\phantom{mn}pq} h_{i\phantom{ab} ef\phantom{mn} pq}^{\phantom{i} ab \phantom{ef} mn}
$.}%
\item $\Ss$ is a projection:
$\Ss_{a_1\ldots a_k\phantom{Ab_1\ldots b_k}B}^{\phantom{a_1\ldots a_k}Ab_1\ldots b_k}$ vanishes if $A\in \mathcal{C}$ or $B \in \mathcal{C}$, otherwise
$$\Ss_{a_1\ldots a_kc_1c_2\phantom{c_3c_4b_1\ldots b_kd_1d_2}d_3d_4}^{\phantom{a_1\ldots a_kc_1c_2}c_3c_4b_1\ldots b_kd_1d_2}
= \delta_{c_1}^{d_1}\delta^{c_3}_{d_3}\varepsilon^{c_4e}\varepsilon_{d_4f} \mathbf{S}^{b_1\ldots b_kd_2f}_{a_1\ldots a_kc_2e}
$$
\item Define the prefix $\sharp$ by $f^{A} = g^{\sharp A}$ iff $f^{\alpha} = g^{\alpha}$, ${f_{ij}}^{k\ell} = {g_{ji}}^{\ell k}$. Similar for $g_{\sharp A}$. Define the prefix $\&$ by\footnote{Observe that ${\Ss^A}_B {\Ss^{\sharp B}}_{\sharp C} = {\Ss^{\sharp A}}_{\sharp B} {\Ss^{B}}_{C}$.}
$f^{\& A} = {\Ss^A}_B {\Ss^{\sharp B}}_{\sharp C} f^C$ and $f_{\& C} = f_A {\Ss^A}_B {\Ss^{\sharp B}}_{\sharp C}$.
\item $\Weyl = \big\{\;({f_{ab}}^A)\;\big|\;\Ss_{ab\phantom{Acd}B}^{\phantom{ab}Acd}{f_{cd}}^B = {f_{ab}}^A \text{ and }{f_{ab}}^{\& A} = 0\big\}$,\; $\dim_{\C}\Weyl = 5$.
\end{enumerate}
\subsubsection{$\SL$, unprimed and primed $\mathcal{S}$-indices}\label{sl2csl2c}
The constructions in this paper are invariant under the action\footnote{The `global action'. That is, the $\SL$-matrix is a constant, not a function.} of $\SL$, the
group of complex $2\times 2$ matrices $A$ with $\varepsilon_{ab}{A^a}_m{A^b}_n = \varepsilon_{mn}$, equivalently ${A^a}_m{A^b}_n\varepsilon^{mn} = \varepsilon^{ab}$.
The matrix $A\in \SL$ acts: trivially on $\mathcal{C}$ indices;
as $A$ on unprimed $\mathcal{S}$ indices;
as $\overline{A}$ on primed $\mathcal{S}$ indices\footnote{Example: $f\mapsto Af$ with $(Af)^{ab'\phantom{c}\mu}_{\phantom{ab'}c} = {A^a}_i \overline{{A^{b'}}_{j'}}{(A^{-1})^k}_c f^{ij'\phantom{k}\mu}_{\phantom{ij'}k}$.}.
The basic fields in this paper are
$\varepsilon^{ab}, \varepsilon^{a'b'}, \varepsilon_{ab}, \varepsilon_{a'b'}$ and
\begin{align}\label{fdhskjfdhkfdhf}
\Back_{ab'} && {F_{ab'}}^A && \xi^A &&  {U_{a'b'}}^A
&& {G^{ab'}}_{A'B} && \eta^A &&  {\Ub_{a'b'}}^A && \zeta^A
\end{align}
The placement of primes determines the action of $\SL$ on each of these fields. $\mathcal{S}$-indices internal to $A$ and $A'$ are unprimed or primed as in $mn'rs'$ and $m'nr's$, respectively\footnote{For example,
${F_{ab'k\ell'}}^{mn'}$ and $G^{ab'k'\ell\phantom{m'n}pq'\phantom{rs'}}_{\phantom{ab'k'\ell}m'n\phantom{pq'}rs'}$.}. \emph{Primes are suppressed in this paper.} They can be restored by referring to \eqref{fdhskjfdhkfdhf}.
\vskip 2mm
Let $(m,n)$ be the irreducible representation of $\SL$ with
$\dim_{\C} = (m+1)(n+1)$, equivalent to the action on tensors that are totally symmetric in $m$ unprimed and totally symmetric in $n$ primed $\mathcal{S}$-indices. Then: 
\begin{enumerate}[$\bullet$]
\item $\xi^A$ transforms as $4(0,0) \oplus ((0,0) \oplus (2,0) \oplus (0,2) \oplus (2,2))$.
The condition $\xi^{\&A}=0$ removes $(2,2)$. The additional condition $\overline{\xi^{A}} = \xi^{\sharp A}$ selects a \emph{real} $\SL$-invariant  subspace, $\dim_\R=11$, which can be thought of as $\Lambda$ from
Subsection \ref{sudskkdhfdk}.
\item $({U_{ab}}^A) \in \Weyl$ if and only if $U$ transforms as $(0,4)$.
\end{enumerate}

The constructions in this paper are invariant under simultaneous complex conjugation of all fields in \eqref{fdhskjfdhkfdhf}. This extends the action to $\SL \rtimes_{\varphi} \Z_2$, with complex conjugation $\varphi: \Z_2 \to \mathrm{Aut} (\SL)$. The action on the $\dim_{\R} = 4$ vector space of all $v^{ab}$ with $\overline{v^{ba}} = v^{ab}$ ($a$ unprimed, $b$ primed) leaves  $(v,w) \mapsto -\varepsilon_{ab}\varepsilon_{mn}v^{am}w^{bn}$ invariant, and  yields a homomorphism from $\SL \rtimes_{\varphi} \Z_2$ \emph{onto} the orthochronous Lorentz group $\mathrm{O}^+(1,3)$. Its kernel $\{\pm\mathbbm{1}_2\}$ acts trivially on the fields in \eqref{fdhskjfdhkfdhf}, because they have an even number of $\mathcal{S}$-indices. Thus, $\mathrm{O}^+(1,3)$ acts on \eqref{fdhskjfdhkfdhf}.

\subsection{Linear model problem (LMP) and gauge-fixing} \label{lmpsec}
The constituents of LMP on $\R^4$ are\footnote{All fields are infinitely differentiable.}:
\begin{itemize}
\item[(i)] Four  linearly independent, complex vector fields $\Back_{ab} = {\Back_{ab}}^{\sigma}\tfrac{\p}{\p x^{\sigma}}$, with constant coefficients, and $\overline{\Back_{ba}} = \Back_{ab}$. They are fixed for the discussion\footnote{Example: $\Back_{00} = \p_0 + \p_1$,\, $\Back_{11} = \p_0 - \p_1$,\, $\Back_{01} = \p_2 + i\p_3$,\, $\Back_{10} = \p_2 - i\p_3$.}.
\item[(ii)] The unknown field ${F_{ab}}^A$, subject to\footnote{${F_{ab}}^A$ is a field on $\R^4$ with values in a $\dim_{\R} = 44$ vector space.} $\overline{{F_{ab}}^{A}} = {F_{ba}}^{\sharp A}$ and ${F_{ab}}^{\&A} = 0$.
\item[(iii)] The field ${U_{mn}}^A = \mathbf{S}_{mn}^{ij}\varepsilon^{ab}\Back_{ai}({F_{bj}}^A)$ associated to the unknown $F$.
\end{itemize}
It follows that:
\begin{itemize}
\item[(iv)] $U$ is invariant under \emph{local gauge transformations} ${F_{ab}}^A \mapsto {F_{ab}}^A + \Back_{ab}(\xi^A)$,
 for any $\xi$ with $\overline{\xi^A} = \xi^{\sharp A}$ and $\xi^{\& A} = 0$, abbreviated
$F \mapsto F + \Back(\xi)$.
\item[(v)] $\varepsilon^{mn}{U_{mn}}^A = 0$ and ${U_{mn}}^{\& A} = 0$.
\item[(vi)] $\overline{\varepsilon^{mn}\Back_{im}({U_{nj}}^A)} - \varepsilon^{mn}\Back_{jm}({U_{ni}}^{\sharp A}) = 0$, differential identities\footnote{Derivation: $\varepsilon^{mn}\Back_{im}({U_{nj}}^A)  = \varepsilon^{mn}\varepsilon^{pq}[\Back_{im}(\Back_{pj}({F_{qn}}^A)) - \tfrac{1}{2} \Back_{ij}(\Back_{pm}({F_{qn}}^A))]$ by (iii) and the identity $\mathbf{S}^{ij}_{ab} =\mathbf{S}^{ij}_{ba} = \delta^i_a\delta^j_b -\tfrac{1}{2}\varepsilon^{ij}\varepsilon_{ab}$.  Now use (i), (ii) and $[\Back_{ab},\Back_{cd}]=0$.} for $U$.
\end{itemize}
By definition, $F$ is a solution to LMP if and only if it solves:
\begin{itemize}
\item[(vii)] Partial differential field equations: $U \in \Weyl$, pointwise.
\end{itemize}
\begin{defn}[Gauge-fixing] Suppose ${G^{ab}}_{AB}$ is constant and:
\begin{itemize}
\item[(G1)] (Hermitian) ${G^{ab}}_{A\&B} = 0$ and $\overline{{G^{ba}}_{BA}} = {G^{ab}}_{AB}$.
\item[(G2)] (Main algebraic condition) ${G^{ab}}_{AB}\Ss_{b\phantom{Bc}C}^{\phantom{b}Bc} = 0$.
\item[(G3)] (Positivity) $\overline{{v_a}^A}{G^{ab}}_{AB}{v_b}^B \geq 0$ for all $v$.\\
Assuming ${v_b}^{\& B} = 0$, equality holds if and only if $\Ss_{a\phantom{Ab}B}^{\phantom{a}Ab}{v_b}^B = {v_a}^A$.
\end{itemize}
A field $F$ as in (ii) is gauge-fixed with respect to $G$ if and only if, pointwise,
\begin{equation}\label{eq:gaugefix}
\tag{$\ast$}
\boxed{\forall \text{$\eta$ with $\overline{\eta^A} = \eta^{\sharp A}$ and $\eta^{\& A}=0$}:\quad
\RE \big(\,\overline{\eta^A} {G^{ab}}_{AB} {F_{ab}}^B\,\big) = 0}
\end{equation}
\end{defn}
Gauge-fixing \eqref{eq:gaugefix} leads to three symmetric hyperbolic systems, introduced in the next sections.
The field ${G^{ab}}_{AB}$ is fixed for the discussion. 
One can read (G4), (G5), (G6) only when they are referred to, later on.
\begin{lemma}\rule{0pt}{0pt}
\begin{itemize}
\item[(G4)] $\overline{{w_a}^A}{G^{ab}}_{AB}{w_b}^B > 0$ if ${w_a}^A = s_a\eta^A$,\; $\eta \neq 0$,\; $\eta^{\& A} = 0$,\; $\overline{\eta^A} = \eta^{\sharp A}$,\; $s\neq 0$.
\end{itemize}
If $t^{ab}$ satisfies $\overline{t^{ba}} = t^{ab}$ and
$\overline{s_a}t^{ab}s_b > 0$ for all $s\neq 0$, then:
\begin{itemize}
\item[(G5)] $\overline{{w_{ma}}^A} (t^{mn}{G^{ab}}_{AB}){w_{nb}}^B > 0$ if $w \neq 0$,\; ${w_{ab}}^{\& A} = 0$,\; $\overline{{w_{ab}}^A} = {w_{ba}}^{\sharp A}$.
\item[(G6)] $\overline{{w_{ma}}^A} (t^{mn}{G^{ab}}_{AB}){w_{nb}}^B \geq 0$ if ${w_{ab}}^{\& A} = 0$,\; $\overline{\varepsilon^{ab}{w_{ab}}^A} = \varepsilon^{ab}{w_{ab}}^{\sharp A}$.\\
Equality if and only if $w \in \Weyl$.
\end{itemize}
\end{lemma}
\emph{Proof.} In (G4), (G5), (G6): ${w_{\ast a}}^{\& A} = 0$, where $\ast$ is nothing or one $\mathcal{S}$-index. By (G3), the bilinear expression is $\geq 0$, with equality iff\footnote{To see this for (G5), (G6), pick $e_a$, $f_a$ with $\overline{e_a}t^{ab}e_b = \overline{f_a}t^{ab}f_b = 1$ and $\overline{e_a}t^{ab}f_b = 0$. Then ${w_{na}}^A = e_n{p_a}^A + f_n{q_a}^A$, with ${p_a}^{\& A} = 0$ and ${q_a}^{\& A} = 0$.} $\Ss_{a\phantom{Ab}B}^{\phantom{a}Ab}{w_{\ast b}}^B = {w_{\ast a}}^A$.
Assume equality. Then $\Ss_{\phantom{A}B}^{\phantom{}A}{w_{\ast a}}^B = {w_{\ast a}}^A$. Thus in (G4), $\Ss_{\phantom{A}B}^{\phantom{}A}\eta^B = \eta^A$, in turn ${\Ss^{\sharp A}}_{\sharp B}\eta^{B} = \eta^{A}$, that is $\eta^A = \eta^{\& A} = 0$, contradicting $\eta \neq 0$. In (G5), $\Ss_{\phantom{A}B}^{\phantom{}A}{w_{ab}}^B = {w_{ab}}^A$,\; ${\Ss^{\sharp A}}_{\sharp B}{w_{ab}}^{B} = {w_{ab}}^{A}$, that is ${w_{ab}}^{A} = {w_{ab}}^{\& A} = 0$, contradicting $w \neq 0$.
In (G6), $\Ss_{\phantom{A}B}^{\phantom{}A}{w_{ab}}^B = {w_{ab}}^A$ and ${w_{ab}}^{\& A}=0$ imply 
${\Ss^{\sharp A}}_{\sharp B}{w_{ab}}^{B} = 0$. Multiply with $\varepsilon^{ab}$, conjugate, find $\varepsilon^{ab}{w_{ab}}^A = 0$. To summarize, ${w_{ab}}^A = {w_{ba}}^A$ and $\Ss_{a\phantom{Ab}B}^{\phantom{a}Ab}{w_{k b}}^B = {w_{k a}}^A$. The group of permutations of $\{1,2,3,4\}$ is generated by the transposition $(12)$ and the six stabilizers of $1$. Thus, in (G6), equality implies $\Ss_{ab\phantom{Acd}B}^{\phantom{ab}Acd}{w_{cd}}^B = {w_{ab}}^A$. The converse holds by (G2). \hfill $\Box$



\subsubsection{First system: the gauge is locally realizable}\label{sec111111}
Assume $F$ as in (ii) is given. Do not assume \eqref{eq:gaugefix} or (vii). The condition for $F + \Back(\xi)$ in (iv) to satisfy \eqref{eq:gaugefix} is a partial differential equation for $\xi$:
\begin{equation}\label{eq:locrea}
\boxed{\forall \text{$\eta$ as in \eqref{eq:gaugefix}:}\qquad 
\RE \big(\,\overline{\eta^A} {G^{ab}}_{AB} \Back_{ab}(\xi^B)\,\big)
= -\RE \big(\,\overline{\eta^A} {G^{ab}}_{AB} {F_{ab}}^B\,\big)}
\end{equation}
This is a symmetric hyperbolic system for $\xi$ ($\R$-dimension $11\times 11$), by (G4).
\subsubsection{Second system: dynamical subset of the field equations}\label{secssec}
From now on, condition \eqref{eq:gaugefix} is assumed.
By (G4), the $\R$-linear functional  associated to any $\eta\neq 0$ in \eqref{eq:gaugefix} is non-trivial.
Thus, \eqref{eq:gaugefix} restricts $F$ to a vector space with $\dim_{\R} = 44-11=33$, that depends only on $G$. Fix a (constant, $\R$-linear) parametrization $\Phi \mapsto F = P(\Phi)$, with $\R^{33}$-valued $\Phi$.
For $P(\Phi)$ to solve the field equations (vii), it is \emph{necessary} that
\begin{subequations}
\begin{equation}\label{sys2}
\boxed{\forall \text{$\Psi$ real:}
\qquad
\RE\Big(\,\overline{{P(\Psi)_{q a}}^A}{G^{ab}}_{AB}\varepsilon^{pq}\varepsilon^{mn}\Back_{m p}\big({P(\Phi)_{n b}}^B\big)\,\Big) = 0}
\end{equation}
or equivalently\footnote{
Replace 
$\Back_{m p}({P(\Phi)_{n b}}^B)
= \delta_p^i\delta_b^j\Back_{m i}({P(\Phi)_{n j}}^B)
= (\mathbf{S}^{ij}_{pb} + \tfrac{1}{2}\varepsilon^{ij}\varepsilon_{pb})\Back_{m i}({P(\Phi)_{n j}}^B)$
in \eqref{sys2}.
Then $\mathbf{S}^{ij}_{pb}$ yields \eqref{sys2b}, but
$\tfrac{1}{2}\varepsilon^{ij}\varepsilon_{pb}$ contributes nothing, because $P(\Psi)$ satisfies \eqref{eq:gaugefix}.
}
\begin{equation}\label{sys2b}
\forall \text{$\Psi$ real:}
\qquad
\RE\Big(\,\overline{{P(\Psi)_{qa}}^A}{G^{ab}}_{AB}\varepsilon^{pq}
{U_{pb}}^B\,\Big) = 0
\end{equation}
\end{subequations}
because ${G^{ab}}_{AB} {U_{pb}}^B = {G^{ab}}_{AB} \Ss_{pb\phantom{Bk\ell}C}^{\phantom{pb}Bk\ell}{U_{k\ell}}^C = 0$, by (vii) and (G2).
Equation \eqref{sys2} is a symmetric hyperbolic system for $\Phi$ ($\R$-dimension $33\times 33$), by (G5)\footnote{As an aside,
consider (G5) with $t^{mn} = \overline{s^m}s^n$, $s\neq 0$, and suppose ${w_{ab}}^A$ satisfies \eqref{eq:gaugefix}.
To analyze equality in (G5) in this \emph{degenerate} case, pick $r^a$ with $\varepsilon_{ab}s^ar^b = 1$.
Set $s_a = \varepsilon_{ab}s^a$, $r_a = \varepsilon_{ab}r^b$.
Then $\delta_a^b = r_as^b - s_ar^b$.
Expand ${w_{ab}}^A = s_a\overline{s_b} \alpha^A + s_a\overline{r_b}\beta^A
+ r_a\overline{s_b} \gamma^A + r_a\overline{r_b}\delta^A$
with $\alpha^{\& A} = \beta^{\& A} = \gamma^{\& A} = \delta^{\& A} = 0$ and
$\overline{\alpha^A} = \alpha^{\sharp A}$, $\overline{\gamma^A} = \beta^{\sharp A}$, $\overline{\delta^A} = \delta^{\sharp A}$.
Equality implies $\Ss_{a\phantom{Ab}B}^{\phantom{a}Ab}s^n{w_{nb}}^B = s^n{w_{na}}^A$ by (G3), implies
$\delta^A = 0$, $\gamma^{\sigma} = 0$, ${\gamma_{am}}^{bn} = z \delta_a^b\overline{s_m s^n}$, $\beta^A = \overline{\gamma^{\sharp A}}$, $z\in \C$; now \eqref{eq:gaugefix} determines $\alpha^A$, see (G4). Conversely, they imply equality, for all $z \in \C$.}.
\subsubsection{Third system: the constraints propagate}
Fix a solution $\Phi$ to \eqref{sys2}, equivalently \eqref{sys2b}.
Here and below, $U$ is the field associated to $P(\Phi)$ through (iii).
Recall $\varepsilon^{ab}{U_{ab}}^A = 0$ from (v). Set
\begin{equation}\label{eehkrhkrehrk}
{\Ub_{ab}}^A = {U_{ab}}^A + \varepsilon_{ab}\zeta^A
\end{equation}
with $\overline{\zeta^A} = \zeta^{\sharp A}$ and $\zeta^{\& A} = 0$.
Irrespective of the eventual choice of $\zeta$:
\begin{itemize}
\item[(v')] $\overline{\varepsilon^{ab}{\Ub_{ab}}^A} = \varepsilon^{ab}{\Ub_{ab}}^{\sharp A}$ and ${\Ub_{ab}}^{\& A} = 0$ by (v).
\item[(vi')] $\overline{\varepsilon^{mn}\Back_{im}({\Ub_{nj}}^A)} - \varepsilon^{mn}\Back_{jm}({\Ub_{ni}}^{\sharp A}) = 0$ by (vi).
\end{itemize}
In addition, \eqref{sys2b} holds with $U$ replaced by $\Ub$, because $P(\Psi)$ satisfies \eqref{eq:gaugefix}. Uniquely fix $\zeta$ by strengthening to:
\begin{equation}\label{sys2c}
\tag{$\ast\ast$}
\forall \text{$F$ as in\footnotemark\,(ii):}
\quad
\RE\Big(\,\overline{{F_{q a}}^A}{G^{ab}}_{AB}\varepsilon^{pq}
{\Ub_{pb}}^B\,\Big) = 0
\end{equation}
\footnotetext{$F$ is a dummy, completely unrelated to $P(\Phi)$, and \emph{does not} have to satisfy \eqref{eq:gaugefix}.}%
Equation \eqref{sys2c} is an $\R$-linear map\footnotemark\,$\text{($U$ satisfying \eqref{sys2b})} \mapsto \zeta$, that depends \emph{only} on $G$. The kernel of this map contains $\Weyl$. Therefore:
\begin{itemize}
\item[(vii')] $\Ub \in \Weyl$ is equivalent to the field equations $U \in \Weyl$ in (vii).
\end{itemize}
\footnotetext{Given a $U$ that satisfies \eqref{sys2b}, interpret \eqref{sys2c} as 44 $\R$-linear equations for $\zeta$.
The 33 coming from $F \in \image P$ hold by \eqref{sys2b}. Denote by $[F] = F + \image P$ elements of the quotient. The main observation is that $([F],\zeta) \mapsto \RE(\,\overline{F_{ba}^{\phantom{ab}A}}{G^{ab}}_{AB}\zeta^B\,)$ is a well defined $\R$-bilinear pairing between vector spaces of equal $\dim_\R = 11$, \emph{non-degenerate} by (G4).}%
$\Ub$ takes values in a vector space of $\dim_\R = 33$,
by (v') and \eqref{sys2c}.
Recall that $\dim_{\R}\Weyl = 10$. Fix a (constant, $\R$-linear) parametrization $\widehat{\Phi} \mapsto \Ub + \Weyl = \Pb(\widehat{\Phi})$ of the $\dim_{\R} = 23$ quotient, with $\R^{23}$-valued $\widehat{\Phi}$. Then 
\begin{subequations}
\begin{equation}\label{sys3}
\boxed{\forall \text{$\widehat{\Psi}$ real:} \qquad \RE\Big(\,\overline{\Pb(\widehat{\Psi})_{qa}^{\phantom{qa}A}} {G^{ab}}_{AB} \varepsilon^{pq}\varepsilon^{mn}\Back_{pm}({\Pb(\widehat{\Phi})_{nb}}^B) \,\Big) = 0}
\end{equation}
To see this, first observe that the left hand side of \eqref{sys3} is well defined by (G2): adding $\Weyl$-valued fields to $\Pb(\widehat{\Psi})$ or $\Pb(\widehat{\Phi})$ does not affect the outcome.
Thus, $\Pb(\widehat{\Phi})$ can be replaced by $\Ub$.
Equation \eqref{sys3} is equivalent to\footnote{The expression $\RE(\overline{\Pb(\widehat{\Psi})_{qa}^{\phantom{qa}A}} {G^{ab}}_{AB} \varepsilon^{pq}{X_{pb}}^B)$ with ${X_{ab}}^{\& A}=0$ is invariant under
${X_{pb}}^B \leadsto - \overline{{X_{bp}}^{\sharp B}}$, because $\Pb(\widehat{\Psi})$ satisfies \eqref{sys2c}, in the role of $\Ub$.
The arrow $\leadsto$ is used again later.}
\begin{equation} \label{dshdkhdkjh}
\forall \text{$\widehat{\Psi}$ real:} \quad \RE\Big(\,\overline{\Pb(\widehat{\Psi})_{qa}^{\phantom{qa}A}} {G^{ab}}_{AB} \varepsilon^{pq}
\Big\{\,\varepsilon^{mn}\Back_{pm}({\Ub_{nb}}^B) - \overline{\varepsilon^{mn}\Back_{bm}({\Ub_{np}}^{\sharp B})}\,\Big\}
\,\Big) = 0
\end{equation}
\end{subequations}
But \eqref{dshdkhdkjh} holds, by (vi'). Therefore, \eqref{sys3} holds as well: a symmetric hyperbolic system for $\widehat{\Phi}$ ($\R$-dimension $23\times 23$), by (G6).
\vskip 4mm
For later reference, record for both
$\bigstar = \delta_p^{\ell} \Ss_{b\phantom{Bc}C}^{\phantom{b}Bc}$
and
$\bigstar = \delta_b^c \Ss_{p\phantom{\sharp B \ell}\sharp C}^{\phantom{p}\sharp B\ell}$,
and for all $Y$ with ${Y_{ab}}^{\& A} = 0$, the identity\footnote{For the first $\bigstar$ identity use (G2). For the second, use (G2) and $\leadsto$ from a previous footnote:
$-(\delta_p^{\ell} \Ss_{b\phantom{Bc}C}^{\phantom{b}Bc}) \overline{{Y_{c \ell}}^{\sharp C}} \leadsto
 \overline{(\delta_b^{\ell} \Ss_{p\phantom{\sharp Bc}C}^{\phantom{p}\sharp Bc}) \overline{{Y_{c \ell}}^{\sharp C}}}
=  (\delta_b^{\ell} \Ss_{p\phantom{\sharp Bc}C}^{\phantom{p}\sharp Bc}){Y_{c \ell}}^{\sharp C}
=  (\delta_b^{c} \Ss_{p\phantom{\sharp B\ell}\sharp C}^{\phantom{p}\sharp B\ell}){Y_{\ell c}}^C$.}
\begin{equation}\label{ehkfdkjdfhldfhfdkha}
\forall \text{$\widehat{\Psi}$ real:} \qquad \RE\big(\,\overline{\Pb(\widehat{\Psi})_{qa}^{\phantom{qa}A}} {G^{ab}}_{AB} \varepsilon^{pq}
\;\bigstar\;
{Y_{\ell c}}^C
\,\big) = 0
\end{equation}
\subsection{From LMP to general relativity (GR)}\label{appgr}
This section is logically organized as follows. 
Subsections \ref{eqdjkhd} and \ref{dshkhfdkjdfjgd} gently replace LMP by GR; the field equations are now quasilinear and contain non-principal terms. Subsection \ref{fddskhkfdshfdhfdk} introduces a more general, inhomogeneous gauge condition. In the light of these modifications, Subsections \ref{localreal}, \ref{revsec}, \ref{revthird} revisit the derivation of  symmetric hyperbolic systems. 

\subsubsection{Translation of LMP to fields with `real four-indices'} \label{eqdjkhd}
The $\dim_{\R} = 4$ vector space of all $v^{ab}$ with $\overline{v^{ba}}=v^{ab}$ ($a$ unprimed, $b$ primed), equipped with $(v,w)\mapsto -\varepsilon_{ab}\varepsilon_{mn}v^{am}w^{bn}$, is $(-,+,+,+)$ Minkowski space. \emph{Thus, each Hermitian unprimed-primed $\mathcal{S}$-index pair is one `real four-index'}.
Below, such pairs are sometimes bracketed, $(am), (bn), \ldots$ When space is short, boldface $\mathbf{a}, \mathbf{b},\ldots$ are used as placeholders for pairs. Set $k_{(am)(bn)} = -\varepsilon_{ab}\varepsilon_{mn}$. Observe that, for any $f$, the condition $f^{\& A} = 0$ is equivalent to
$$ {f_{(am)}}^{(ck)}k_{(ck)(bn)} + {f_{(bn)}}^{(ck)}k_{(ck)(am)} - \tfrac{1}{2}{f_{(ck)}}^{(ck)}k_{(am)(bn)} = 0
$$
\vskip 1mm
Items (i), (ii) require no translation. To translate (iii), (v), (vi), (vii), a new field ${U_{(am)(bn)}}^A$ is introduced instead of ${U_{mn}}^A$:
\begin{itemize}
\item[(iii'')] ${U_{(am)(bn)}}^A = \Back_{(am)}({F_{(bn)}}^A)-\Back_{(bn)}({F_{(am)}}^A)$, equivalent\footnote{In fact, ${U_{mn}}^A = \tfrac{1}{2}\varepsilon^{ab}{U_{ambn}}^A$ and ${U_{ambn}}^A = \varepsilon_{ab}{U_{mn}}^A + \varepsilon_{mn}\overline{{U_{ba}}^{\sharp A}}$. For the second, write ${U_{ambn}}^A = 
(\mathbf{S}_{ab}^{ij}+\tfrac{1}{2}\varepsilon_{ab}\varepsilon^{ij})
(\mathbf{S}_{mn}^{pq}+\tfrac{1}{2}\varepsilon_{mn}\varepsilon^{pq})
{U_{ipjq}}^A$, then exploit ${U_{ambn}}^A + {U_{bnam}}^A = 0$.
} to
${U_{mn}}^A$.
\item[(v'')] ${U_{(am)(bn)}}^A = - {U_{(bn)(am)}}^A$,\;
${U_{(am)(bn)}}^{\& A} = 0$ and 
$\overline{{U_{(am)(bn)}}^A} = {U_{(ma)(nb)}}^{\sharp A}$.
\item[(vi'')] $\Back_{(am)}({U_{(bn)(ck)}}^A)+\Back_{(bn)}({U_{(ck)(am)}}^A)+\Back_{(ck)}({U_{(am)(bn)}}^A)=0$.
\item[(vii'')] $({U_{(am)(bn)}}^A) \in \Weyl_{\R}$.
\end{itemize}
Here, $\Weyl_{\R}$ is the vector space of all $({f_{(am)(bn)}}^A)$ with 
$\overline{{f_{(am)(bn)}}^A} = {f_{(ma)(nb)}}^{\sharp A}$
and
${f_{\mathbf{a}\mathbf{b}}}^A = -{f_{\mathbf{b}\mathbf{a}}}^A$
and
${f_{\mathbf{a}\mathbf{b}}}^{\& A} = 0$
and ${f_{\mathbf{a}\mathbf{b}}}^{\sigma} = 0$
and ${f_{\mathbf{i}\mathbf{j}\mathbf{k}}}^{\mathbf{a}}
+ {f_{\mathbf{j}\mathbf{k}\mathbf{i}}}^{\mathbf{a}}
+ {f_{\mathbf{k}\mathbf{i}\mathbf{j}}}^{\mathbf{a}}=0$
and ${f_{\mathbf{i}\mathbf{k}\mathbf{a}}}^{\mathbf{a}} = 0$
and ${f_{\mathbf{i}\mathbf{a}\mathbf{k}}}^{\mathbf{a}} = 0$. Observe that 
$\dim_{\R}\Weyl_{\R} = 10$.
\vskip 3mm
Translations as in this subsection are implicit in the discussion below.
\subsubsection{Orthonormal frame formalism for GR}\label{dshkhfdkjdfjgd}
\newcommand{\B}{\mathbf}
\newcommand{\Bl}{\boldsymbol{\ell}}
The discussion about GR is local on $\R^4$. Set $\text{\GR{ii}} = \text{(ii)}$, 
$\text{\GR{v}} = \text{(v'')}$. The first implies, in particular, ${F_{ab}}^{\sigma} = \overline{{F_{ba}}^{\sigma}}$.
\begin{itemize}
\item[\GR{i}] $F_{(ab)} = {F_{(ab)}}^{\sigma}\tfrac{\p}{\p x^{\sigma}}$ are linearly independent (`non-degenerate frame').
\item[\GR{iii}] Set ${U_{(am)(bn)}}^A = F_{(am)}({F_{(bn)}}^A)-F_{(bn)}({F_{(am)}}^A) + p(F,F)$ for a polynomial $p$. Explicitly, with $\mathbf{A}^{\mathbf{a}_1\ldots \mathbf{a}_k}_{\mathbf{b}_1\ldots \mathbf{b}_k}
= \sum_{\pi \in S_k} \mathrm{sgn}(\pi)\, \delta^{\mathbf{a}_1}_{\mathbf{b}_{\pi(1)}}\cdots \delta^{\mathbf{a}_k}_{\mathbf{b}_{\pi(k)}}$:
\begin{align*}
{U_{\B{i}\B{j}}}^{\sigma} &= \B{A}_{\B{i}\B{j}}^{\B{b}\B{c}}\big(F_{\B{b}}({F_{\B{c}}}^{\sigma})
- {F_{\B{b}\B{c}}}^{\Bl} {F_{\Bl}}^{\sigma}\big)\\
{U_{\B{i}\B{j}\B{m}}}^{\B{n}} & = \B{A}_{\B{i}\B{j}}^{\B{b}\B{c}}\big(
F_{\B{b}}({F_{\B{c}\B{m}}}^{\B{n}}) + {F_{\B{c}\B{m}}}^{\Bl}{F_{\B{b}\Bl}}^{\B{n}}
- {F_{\B{b}\B{c}}}^{\Bl}{F_{\Bl \B{m}}}^{\B{n}}\big)
\end{align*}
\item[\GR{iv}] For all $(\varphi,\Delta)$,  with $\varphi$ a diffeomorphism (of open subsets of $\R^4$) and  ${\Delta^{(am)}}_{(bn)} = \Theta^2 {A^a}_b\overline{{A^m}_n}$ (where the field $A$ is $\SL$-valued and the field $\Theta > 0$), the local gauge transformation $F \mapsto \widetilde{F}$ given by
\begin{align*}
{\widetilde{F}_{\mathbf{i}}}^{\phantom{\mathbf{i}}\sigma} \circ \varphi & = {\Delta^{\mathbf{a}}}_{\mathbf{i}}\, {F_{\mathbf{a}}}(\varphi^{\sigma})\\
{\widetilde{F}_{\mathbf{i}\mathbf{m}}}^{\phantom{\mathbf{i}\mathbf{m}}\mathbf{n}} \circ \varphi &  =
{\Delta^{\mathbf{a}}}_{\mathbf{i}}\Big(
{F_{\mathbf{a}\mathbf{k}}}^{\boldsymbol{\ell}}  {\Delta^{\mathbf{k}}}_{\mathbf{m}} {(\Delta^{-1})^{\mathbf{n}}}_{\boldsymbol{\ell}} + {(\Delta^{-1})^{\mathbf{n}}}_{\boldsymbol{\ell}} {F_{\mathbf{a}}}({\Delta^{\boldsymbol{\ell}}}_{\mathbf{m}})\Big)
\end{align*}
implies the transformation $U \mapsto \widetilde{U}$ given by, with $U_{\B{a}\B{b}} = {U_{\B{a}\B{b}}}^{\sigma} \tfrac{\p}{\p x^{\sigma}}$:
\begin{align*}
{\widetilde{U}_{\mathbf{i}\mathbf{j}}}^{\phantom{\mathbf{i}\mathbf{j}}\sigma} \circ \varphi & =
{\Delta^{\mathbf{a}}}_{\mathbf{i}}
{\Delta^{\mathbf{b}}}_{\mathbf{j}}
\, {U_{\mathbf{a}\mathbf{b}}}(\varphi^{\sigma})\\
{\widetilde{U}_{\mathbf{i}\mathbf{j}\mathbf{m}}}^{\phantom{\mathbf{i}\mathbf{j}\mathbf{m}}\mathbf{n}} \circ \varphi &  =
{\Delta^{\mathbf{a}}}_{\mathbf{i}}
{\Delta^{\mathbf{b}}}_{\mathbf{j}}
\Big({U_{\mathbf{a}\mathbf{b}\mathbf{k}}}^{\boldsymbol{\ell}}  {\Delta^{\mathbf{k}}}_{\mathbf{m}} {(\Delta^{-1})^{\mathbf{n}}}_{\boldsymbol{\ell}} + {(\Delta^{-1})^{\mathbf{n}}}_{\boldsymbol{\ell}} {U_{\mathbf{a}\mathbf{b}}}({\Delta^{\boldsymbol{\ell}}}_{\mathbf{m}})\Big)
\end{align*}
\item[\GR{vi}] $F_{(am)}({U_{(bn)(ck)}}^A)+F_{(bn)}({U_{(ck)(am)}}^A)+F_{(ck)}({U_{(am)(bn)}}^A) + q(F\oplus \p F, U) =0$, with $q$ a polynomial, are differential identities for $U$.
Explicitly:
\begin{subequations}
\begin{align}
\label{dfhkhdkhfdkA}
0 & = \tfrac{1}{2} \B{A}_{\B{i}\B{j}\B{k}}^{\B{b}\B{c}\B{d}}\big(
F_{\B{b}}({U_{\B{c}\B{d}}}^{\sigma}) - U_{\B{c}\B{d}}({F_{\B{b}}}^{\sigma})
- 2 {F_{\B{b}\B{c}}}^{\Bl} {U_{\Bl \B{d}}}^{\sigma}
+ {U_{\B{c}\B{d}\B{b}}}^{\Bl} {F_{\Bl}}^{\sigma} \big)\\
\notag
0 & = \tfrac{1}{2} \B{A}_{\B{i}\B{j}\B{k}}^{\B{b}\B{c}\B{d}}\big(
F_{\B{b}}({U_{\B{c}\B{d}\B{m}}}^{\B{n}})-
U_{\B{c}\B{d}}({F_{\B{b}\B{m}}}^{\B{n}})
+ {U_{\B{c}\B{d}\B{m}}}^{\Bl}{F_{\B{b}\Bl}}^{\B{n}}\\
\label{dfhkhdkhfdkB}
& \hskip 76pt - {F_{\B{b}\B{m}}}^{\Bl} {U_{\B{c}\B{d}\Bl}}^{\B{n}}
- 2 {F_{\B{b}\B{c}}}^{\Bl} {U_{\Bl \B{d}\B{m}}}^{\B{n}}
+ {U_{\B{c}\B{d}\B{b}}}^{\Bl}{F_{\Bl\B{m}}}^{\B{n}}
\big)
\end{align}
\end{subequations}
\item[\GR{vii}] Field equations: $({U_{(am)(bn)}}^A) \in \Weyl_{\R}$.\\
Note that gauge transformations \GR{iv} map solutions to solutions.
\end{itemize}
These formulas are derived in \cite{RT}. In the notation of \cite{RT}, \GR{ii}: $\Diamond^1 \in \mathcal{P}^1$. \GR{iii}: $\Diamond^2 = \tfrac{1}{2}\llbracket \Diamond^1,\Diamond^1\rrbracket  \in \mathcal{P}^2$ where $\llbracket\,\cdot\,,\,\cdot\,\rrbracket: \mathcal{P}^k\times \mathcal{P}^{\ell} \to \mathcal{P}^{k+\ell}$ is a Lie superbracket.
\GR{iv}:  equation (7.8) \cite{RT}.
\GR{vi}: $\llbracket \Diamond^1, \Diamond^2\rrbracket = 0$ by the super Jacobi identity. \GR{vii}: $\Diamond^2 \in \mathcal{P}_{\text{vac}}^2$ where $\mathcal{P}_{\text{vac}}^2 \subset \mathcal{P}^2$ is the Weyl sector.
See Proposition 4.5 \cite{RT}.  $F$ and $U$ are the components of $\Diamond^1$ and $\Diamond^2$.

This formalism is a subformalism of, and equivalent to\footnote{The solution spaces are one-to-one, assuming sufficient differentiability of the solutions.}, the Newman-Penrose-Friedrich \cite{NP}, \cite{Fr} formalism, with ${F_{(ab)}}^A$ the frame ($A\in \mathcal{C}$) and connection ($A\notin \mathcal{C}$),
${U_{(am)(bn)}}^A$ the torsion ($A\in \mathcal{C}$) and curvature ($A\notin \mathcal{C}$).
\subsubsection{Gauge-fixing revisited for GR}\label{fddskhkfdshfdhfdk}

In the LMP discussion,
it was assumed that ${G^{ab}}_{AB}$ and the parametrizations $P$ and $\Pb$ are constant. 
\emph{These assumptions are now dropped.} Furthermore, the homogeneous \eqref{eq:gaugefix} is replaced by the inhomogeneous
\begin{equation}\label{eq:gcgr}
\tag{$\ast_{\text{GR}}$}
\boxed{\forall \text{$\eta$ with $\overline{\eta^A} = \eta^{\sharp A}$ and $\eta^{\& A}=0$}:\quad
\RE \big(\,\overline{\eta^A} {G^{ab}}_{AB} {F_{ab}}^B\,\big) = \overline{\eta^A}\,i_A}
\end{equation}
where the field $i_A$, with $i_{\& A} = 0$ and $\overline{i_A} = i_{\sharp A}$, is
 fixed beforehand, like ${G^{ab}}_{AB}$.
The inhomogeneity $i_A$ has $11$ real components, and is important for \GR{i}.

The parametrization becomes $\Phi \mapsto F = F_0 + P(\Phi)$, where $F_0$ is chosen beforehand and satisfies \GR{ii}
and \eqref{eq:gcgr}. On the other hand, $\Phi \mapsto P(\Phi)$ still parametrizes \GR{ii} = (ii) and the homogeneous \eqref{eq:gaugefix}.

\subsubsection{First system revisited: the gauge is locally realizable in GR} \label{localreal}
Assume $F$ as in \GR{ii} is given. Don't assume \eqref{eq:gcgr} or \GR{vii}. The condition for $\widetilde{F}$ in \GR{iv} to satisfy \eqref{eq:gcgr} is a partial differential equation for $(\varphi,\Delta)$:
\begin{align}\label{fdkhkdhkfdhdfkhfdkhdk}
&\forall \text{$\eta$ as in \eqref{eq:gcgr}}:&
\RE \big(\,\overline{\eta^A} ({G^{ab}}_{AB}\circ \varphi) \;\bigstar\big) & = \overline{\eta^A}(i_A\circ \varphi)
\end{align}
where $\bigstar = \widetilde{F}_{ab}^{\phantom{ab}B}\circ \varphi$
is expressed in terms of $F$, $\varphi$, $\Delta$ using \GR{iv}. The principal part consists only of the terms involving
$F_{\mathbf{a}}(\varphi^{\sigma})$ and
$F_{\mathbf{a}}({\Delta^{\boldsymbol{\ell}}}_{\mathbf{m}})$. Parametrize\footnote{A local parametrization around ${\xi_{\mathbf{m}}}^{\mathbf{n}} = 0$ suffices for this subsection.}
$$\varphi^{\sigma} = \xi^{\sigma}\qquad 
{\Delta^{\mathbf{a}}}_{\mathbf{b}} = {\exp(\xi)^{\mathbf{a}}}_{\mathbf{b}} = \textstyle\sum_{k=0}^{\infty} \frac{1}{k!} {\xi_{\mathbf{b}}}^{\mathbf{n}_1}
{\xi_{\mathbf{n}_1}}^{\mathbf{n}_2} \cdots {\xi_{\mathbf{n}_{k-1}}}^{\mathbf{a}}$$
where $\overline{\xi^{A}} = \xi^{\sharp A}$ and $\xi^{\& A} = 0$. 
For every $\xi$, the map $\kappa \mapsto \mathcal{D}(\xi,\kappa)$ given by $$\mathcal{D}(\xi,\kappa)^{\sigma} = \kappa^{\sigma}\qquad {\mathcal{D}(\xi,\kappa)_{\mathbf{m}}}^{\mathbf{n}} =
{\exp(-\xi)^{\mathbf{n}}}_{\Bl}\;\tfrac{\dd}{\dd s}\big|_{s=0} {\exp(\xi+s\kappa)^{\Bl}}_{\mathbf{m}}
$$
 is an invertible $\R$-linear map on the vector space given by  $\overline{\kappa^{A}} = \kappa^{\sharp A}$, $\kappa^{\& A} = 0$. Therefore, equation \eqref{fdkhkdhkfdhdfkhfdkhdk} is equivalent to
\begin{equation}
\boxed{\forall \text{$\eta$ as in \eqref{eq:gcgr}}:\; 
\RE \big(\,\overline{\mathcal{D}(\xi,\eta)^A} {\Delta^{mn}}_{ab} ({G^{ab}}_{AB}\circ \varphi) {F_{mn}}^{\sigma} \mathcal{D}(\xi,\tfrac{\p}{\p x^{\sigma}}\xi)^B\big) = \text{(npt)}}
\end{equation}
where the non-principal terms \text{(npt)} are \emph{without} derivatives of $\xi$.
This is a symmetric hyperbolic system for $\xi$. It is analogous to \eqref{eq:locrea}, but quasilinear.
\subsubsection{Second system revisited}\label{revsec}
Assume \eqref{eq:gcgr}. Use $\Phi \mapsto F = F_0 + P(\Phi)$ from Subsection \ref{fddskhkfdshfdhfdk}.
Then  ${U_{mn}}^A = \mathbf{S}^{ij}_{mn} \varepsilon^{ab}F_{ai}({P(\Phi)_{bj}}^A) + \text{(npt)}$ by \GR{iii}. The non-principal terms $\text{(npt)}$ are without derivatives of $\Phi$. The field equations \GR{vii} imply \eqref{sys2b}, equivalently
\begin{equation}
\boxed{\forall \text{$\Psi$ real:}
\qquad
\RE\Big(\,\overline{{P(\Psi)_{q a}}^A}{G^{ab}}_{AB}\varepsilon^{pq}\varepsilon^{mn}F_{m p}\big({P(\Phi)_{n b}}^B\big)\,\Big) =
\text{(npt)}}
\end{equation}
analogous to \eqref{sys2}.
The equation is quasilinear, because $F_{mp}$ depends on $\Phi$.
\subsubsection{Third system revisited}\label{revthird}
Fix a solution $\Phi$ to \eqref{sys2b}, where $U$ is the field associated to $F_0 + P(\Phi)$ through \GR{iii}.
Adopt \eqref{eehkrhkrehrk} and \eqref{sys2c} verbatim to define $\Ub$. Then $\text{\GR{v'}} = \text{(v')}$ and
\begin{itemize}
\item[\GR{vi'}] $\overline{\varepsilon^{mn}F_{im}({\Ub_{nj}}^A)} - \varepsilon^{mn}F_{jm}({\Ub_{ni}}^{\sharp A}) = 
\overline{\varepsilon^{mn}F_{im}({U_{nj}}^A)} - \varepsilon^{mn}F_{jm}({U_{ni}}^{\sharp A}) = \text{(npt)}$ by \GR{vi}. The terms $\text{(npt)}$ are without derivatives of $U$ or $\Ub$.
\end{itemize}
and $\text{\GR{vii'}} = \text{(vii')}$. Parametrize $\widehat{\Phi}\mapsto \Ub + \Weyl = \Pb(\widehat{\Phi})$.
Equation \eqref{dshdkhdkjh} holds if $\Back$ is replaced by $F$ and appropriate non-principal terms $\text{(npt)}$ from \GR{vi'} are added. The resulting system is  equivalent to
\begin{equation}\label{erhkhdlshhfdkfdh}
\boxed{\forall \text{$\widehat{\Psi}$ real:} \qquad \RE\Big(\,\overline{\Pb(\widehat{\Psi})_{qa}^{\phantom{qa}A}} {G^{ab}}_{AB} \varepsilon^{pq}\varepsilon^{mn}F_{pm}({\Pb(\widehat{\Phi})_{nb}}^B) \,\Big) = \text{(npt)}}
\end{equation}
analogous to \eqref{sys3}. The non-principal terms $\text{(npt)}$ are without derivatives of, and linear homogeneous in, $U$ or $\Ub$, see \GR{vi} and \GR{vi'}. \emph{However, to use \eqref{erhkhdlshhfdkfdh} to show that the constraints propagate, it is essential that the} $\text{(npt)}$ \emph{in \eqref{erhkhdlshhfdkfdh} descend to linear homogeneous functions of $\Pb(\widehat{\Phi})$}.

To see that this is the case, hypothetically add in \GR{vi} an arbitrary field ${X_{\mathbf{a}\mathbf{b}}}^A$ with values in $\Weyl_{\R}$ to ${U_{\mathbf{a}\mathbf{b}}}^A$.
Now, \eqref{dfhkhdkhfdkA} and \eqref{dfhkhdkhfdkB} are potentially violated. But \eqref{dfhkhdkhfdkA} is not violated, by the definition of $\Weyl_{\R}$. 
The right hand side of \eqref{dfhkhdkhfdkB} can be rewritten
$\tfrac{1}{2} \B{A}_{\B{i}\B{j}\B{k}}^{\B{b}\B{c}\B{d}}{Y_{\mathbf{b}\mathbf{c}\mathbf{d}\mathbf{m}}}^{\mathbf{n}}$
where ${Y_{\mathbf{b}\mathbf{c}\mathbf{d}\mathbf{m}}}^{\mathbf{n}}=
F_{\B{b}}({X_{\B{c}\B{d}\B{m}}}^{\B{n}})
+ {F_{\B{b}\Bl}}^{\B{n}} {X_{\B{c}\B{d}\B{m}}}^{\Bl}
- {F_{\B{b}\B{m}}}^{\Bl} {X_{\B{c}\B{d}\Bl}}^{\B{n}}
- {F_{\B{b}\B{c}}}^{\Bl} {X_{\Bl \B{d}\B{m}}}^{\B{n}}
- {F_{\B{b}\B{d}}}^{\Bl} {X_{\B{c}\Bl\B{m}}}^{\B{n}}$.
For fixed index $\mathbf{b}$, the field ${Y_{\mathbf{b}\mathbf{c}\mathbf{d}\mathbf{m}}}^{\mathbf{n}}$ is in $\Weyl_{\R}$.
Thus, in the notation of Subsection \ref{sl2csl2c}, the field $Y$ is in $(1,1)\otimes_{\C}((4,0)\oplus (0,4)) = (5,1)\oplus (3,1)\oplus (1,3)\oplus (1,5)$. 
In turn, $\tfrac{1}{2} \B{A}_{\B{i}\B{j}\B{k}}^{\B{b}\B{c}\B{d}}{Y_{\mathbf{b}\mathbf{c}\mathbf{d}\mathbf{m}}}^{\mathbf{n}}$, being antisymmetric in $\B{i}\B{j}\B{k}$, is in $(3,1)\oplus (1,3)$. Such violations of \eqref{dfhkhdkhfdkB} do not contribute to \eqref{erhkhdlshhfdkfdh}: those in $(3,1)$ do not contribute by the second $\bigstar$ in \eqref{ehkfdkjdfhldfhfdkha}, those in $(1,3)$ do not contribute by the first $\bigstar$ in \eqref{ehkfdkjdfhldfhfdkha}.

A different proof uses $\llbracket \Diamond^1,\mathcal{P}_{\text{vac}}^2 \rrbracket \subset \mathcal{P}^3_{\text{vac}}$ from Lemma 5.1 \cite{RT}, then \eqref{ehkfdkjdfhldfhfdkha}.
\subsubsection{Systematic choice of ${G^{ab}}_{AB}$} \label{ddd:dkhfkhdfkdf}
Under the $\SL$-action in Subsection \ref{sl2csl2c}, ${G^{ab}}_{AB}$ transforms as a direct sum of irreducible representations, all of type $(1,1)$, $(1,3)$, $(3,1)$, $(3,3)$. The trivial representation $(0,0)$ does not appear. Thus, no nonzero ${G^{ab}}_{AB}$ is $\SL$-invariant.

However, to every subgroup $H \subset \SL$ one can associate the subspace of all  $H$-invariant ${G^{ab}}_{AB}$.
\emph{The positive\footnote{Positivity in the sense of (G3).} elements of this subspace are the natural candidates for ${G^{ab}}_{AB}$, if a given problem or physical situation is symmetric with respect to\footnotemark\:$H$.}
\footnotetext{The discussion is oversimplified. Lorentz transformations that change the orientation, and\,/\,or flip future and past light cones, are ignored. $\mathcal{C}$-indices are ignored. The discussion is point-by-point. The field character of ${G^{ab}}_{AB}$ 
(Subsection \ref{fddskhkfdshfdhfdk}) is ignored. And so forth.}

For example, consider physical situations in which only the timelike vector $\delta^{ab}$ ($a$ unprimed, $b$ primed) is distinguished. The stabilizer group is $\SU \subset \SL$. A particular example of an $\SU$-invariant gauge that satisfies (G1), (G2), (G3) is given by 
${G^{ab}}_{AB} = \delta^{pq}\delta_{PQ} \boldsymbol{\pi}_{p\phantom{Pa}A}^{\phantom{p}Pa}
\boldsymbol{\pi}_{q\phantom{Qb}B}^{\phantom{q}Qb}$, with the projection operator $\boldsymbol{\pi}_{a\phantom{Ab}B}^{\phantom{a}Ab} = (\delta_a^c\delta^A_C - \Ss_{a\phantom{Ac}C}^{\phantom{a}Ac})(\delta_c^b\delta_B^C - \delta_c^b {\Ss^C}_E{\Ss^{\sharp E}}_{\sharp B})$.

\end{document}